\documentclass[12pt]{article}
\usepackage[reqno]{amsmath}
\usepackage{bbm}
\usepackage{epsfig}
\usepackage{array}
\usepackage{float}

\usepackage{a4}

\usepackage{a4wide}
\usepackage{wasysym}

\def\Jo#1#2#3#4{{\it #1} {\bf #2}, #3 (#4)}

\def\NPB{{Nucl. Phys.} {\bf B}}

\def\PLB{{Phys. Lett.}  {\bf B}}
\def\PRL{Phys. Rev. Lett.}
\def\PRD{{Phys. Rev.} {\bf D}}

\def\ZPC{{Z. Phys.} {\bf C}}
\def\IJMP{Int. J. of Mod. Phys. {\bf A}}
\def\JHEP{JHEP}

\def\be{\begin{equation}}
\def\ee{\end{equation}}
\def\gs{\mathrel{
   \rlap{\raise 0.511ex \hbox{$>$}}{\lower 0.511ex \hbox{$\sim$}}}}
\def\ls{\mathrel{
   \rlap{\raise 0.511ex \hbox{$<$}}{\lower 0.511ex \hbox{$\sim$}}}}

\newcommand{\onbb}{neutrinoless double beta decay }
\newcommand{\ba}{\begin{array}{c}}
\newcommand{\baz}{\begin{array}{cc}}
\newcommand{\bad}{\begin{array}{ccc}}
\newcommand{\bea}{\begin{equation} \begin{array}{c}}
\newcommand{\eea}{ \end{array} \end{equation}}
\newcommand{\ea}{\end{array}}
\newcommand{\D}{\displaystyle}
\newcommand{\dms}{\mbox{$\Delta m^2_{\odot}$}}
\newcommand{\dma}{\mbox{$\Delta m^2_{\rm A}$}}
\newcommand{\meff}{\mbox{$\langle m \rangle$}}
\newcommand{\eV}{\mbox{ eV}}
\newcommand{\ppp}{\mbox{$(+++)$ }}
\newcommand{\pmm}{\mbox{$(+--)$ }}
\newcommand{\mpm}{\mbox{$(-+-)$ }}
\newcommand{\mmp}{\mbox{$(--+)$ }}

\begin{document}
\title{
\hfill {\small Ref. SISSA 73/2003/EP}\\
\hfill {\small hep-ph/0308119} \\ \vskip 1cm
\bf Neutrino mass matrices leaving no trace 
}
\author{
W.\ Rodejohann\footnote{Email: {\tt werner@sissa.it}}\\[0.3cm] 
{\normalsize \it Scuola Internazionale Superiore di Studi Avanzati,
I-34014 Trieste, Italy}\\ 
{\normalsize and }\\
{\normalsize \it Istituto Nazionale di Fisica Nucleare,
Sezione di Trieste, I-34014 Trieste, Italy}\\}
\date{}
\maketitle
\thispagestyle{empty}
\begin{abstract}
\noindent We point attention to the fact that in $SO(10)$ models 
with non--canonical (type II) see--saw mechanism 
and exact $b-\tau$ unification the trace of the 
neutrino mass matrix is very small, in fact practically zero. 
This has the advantage of being a basis independent feature. 
Taking a vanishing trace as input, immediate phenomenological 
consequences for the values of the neutrino masses, the $CP$ phases 
or the predictions for neutrinoless double beta decay arise.  
We analyze the impact of the zero trace condition for the normal 
and inverted mass ordering and in case of $CP$ 
conservation and violation. 
Simple candidate mass matrices  
with (close to) vanishing trace and non--zero $U_{e3}$ are proposed. 
We also compare the results with the other most simple 
basis independent property, namely a vanishing determinant.

\end{abstract}

\newpage

\section{\label{sec:intro}Introduction}
The neutrino mass matrix $m_\nu$ contains more parameters than can 
be measured in realistic experiments. This concerns 
in particular the lightest of the three mass eigenstates and 
one if not both of the Majorana phases \cite{ichCP}. In addition, if the 
mixing matrix element $|U_{e3}|$ is too small, also the Dirac phase will 
be unobservable. Thus, the presence of certain 
conditions or simplifications of the neutrino mass matrix is more than 
welcome. What comes first to one's mind is of course  
the presence of zeros in the mass matrix \cite{zeros}. However, zeros in 
a certain basis must not be zeros in another one, so that basis 
independent conditions are advantageous to consider. 
Any matrix possesses two basis independent quantities, namely its 
trace and its determinant. The most simple situation is present if these 
quantities are zero. 
The condition 
$\det m_\nu = 0$ \cite{det}  
leads to one neutrino with vanishing mass and courtesy of 
this fact one gets also rid of one of the notorious Majorana phases. 
A vanishing determinant 
can be motivated on various grounds \cite{detmot1,detmot2}.  
The second, most simple basis independent requirement is a 
vanishing trace, i.e., ${\rm Tr} \, m_\nu = 0$.  
Its consequences have first been investigated in \cite{Tr1} applying a 
three neutrino framework that simultaneously explains the anomalies of 
solar and atmospheric neutrino oscillation 
experiments as well as the LSND experiment. 
In \cite{Tr2}, the $CP$ conserving 
traceless $m_\nu$ has been investigated for the more realistic case 
of explaining only the atmospheric and solar neutrino deficits. 
Motivation of traceless mass matrices can be provided by models in which 
$m_\nu$ is constructed through a commutator of two matrices, as it 
happens in models of radiative mass generation \cite{modrad}. 
More interestingly, and stressed here, a 
(close to) traceless $m_\nu$ can be the consequence of exact $b-\tau$ 
unification at high scale within type II see--saw models \cite{typeII}, 
which in this framework is also the reason for maximal 
atmospheric neutrino mixing \cite{btau1,btau2}. 
The type II see--saw mechanism was the 
original motivation of the traceless $m_\nu$ condition as 
investigated in \cite{Tr1}.\\

\noindent In this letter we shall investigate the outcomes of 
the requirement ${\rm Tr} \, m_\nu = 0$ for the values of the 
neutrino masses and in case of $CP$ violation also of the $CP$ phases. 
We investigate the predictions for observables such as the effective 
mass measured in neutrinoless double beta decay and 
compare them with the ones stemming from the zero determinant case. 
Simple forms of $m_\nu$ that accomplish the 
traceless condition and allow for simple correlations between the 
mixing parameters, mass squared differences and 
the effective mass as measurable 
in neutrinoless double beta decay are presented.

\section{\label{sec:frame}Framework}

\subsection{\label{sec:data}Data}
The light neutrino Majorana mass matrix $m_\nu$ is observable in terms of 
\be \label{eq:mnu}
m_\nu = U \, \, m_\nu^{\rm diag} \, U^T~.
\ee
Here $m_\nu^{\rm diag}$ is a diagonal matrix containing the 
light neutrino mass eigenstates $m_i$. For the normal mass 
ordering (NH) one has $|m_3| > |m_2| > |m_1|$, whereas the inverted mass 
ordering (IH) implies $|m_2| > |m_1| > |m_3|$. 
Mixing is described by $U$, the unitary 
Pontecorvo--Maki--Nagakawa--Sakata \cite{PMNS} lepton 
mixing matrix, which can be parametrized as 
\bea \label{eq:Upara}
U = \left( \bad 
c_{12} c_{13} & s_{12} c_{13} & s_{13} e^{-i \delta} \\[0.2cm] 
-s_{12} c_{23} - c_{12} s_{23} s_{13} e^{i \delta} 
& c_{12} c_{23} - s_{12} s_{23} s_{13} e^{i \delta} 
& s_{23} c_{13} \\[0.2cm] 
s_{12} s_{23} - c_{12} c_{23} s_{13} e^{i \delta} & 
- c_{12} s_{23} - s_{12} c_{23} s_{13} e^{i \delta} 
& c_{23} c_{13}\\ 
               \ea   \right) 
 {\rm diag}(1, e^{i \alpha}, e^{i (\beta + \delta)}) \, , 
\eea
where $c_{ij} = \cos\theta_{ij}$ and $s_{ij} = \sin\theta_{ij}$. 
The phases are usually distinguished as the ``Dirac phase'' $\delta$ 
and the ``Majorana phases'' \cite{STP} $\alpha$ and $\beta$. 
The former can be measured in oscillation experiments, whereas 
the latter show up only in lepton number violating processes. 
Their influence on 
the values of the mass matrix elements is known \cite{ichJPG,mnu}, however, 
only the $ee$ element of $m_\nu$ can realistically be expected 
to be measured \cite{ichPRD,ichJPG}. 
 
\noindent In case of $CP$ 
conservation, different relative signs of the masses $m_i$ are 
possible, corresponding to the intrinsic $CP$ parities of the 
neutrinos \cite{phaalt,ichNPB}. Four situations are possible, 
with $m_i = \eta_i |m_i|$ one can write these cases 
as \ppp$\!\!$, \pmm$\!\!$, \mpm and \mmp$\!\!$, where the 
$(\pm\pm\pm)$ correspond to the relative signs of the mass states. 
Special values of the phases correspond to these 
sign signatures \cite{ichNPB}: 
\be \label{eq:CPV}
\bad (+++) & 
\eta_1 = \eta_2 = \eta_3 = 1 
& \leftrightarrow \alpha = \beta = \pi \\[0.2cm]
     (+--) & 
\eta_1 = - \eta_2 = - \eta_3 = 1 
& \leftrightarrow \alpha = \beta = \frac{\pi}{2}\\[0.2cm]
     (-+-) & 
\eta_1 = - \eta_2 = \eta_3 = -1 
& \leftrightarrow \alpha = \frac{\beta}{2} = \frac{\pi}{2}\\[0.2cm]
     (--+) & 
\eta_1 = \eta_2 = - \eta_3 = -1 
& \leftrightarrow \alpha = 2 \beta = \pi
\ea \, .
\ee
Observation implies the following values of the oscillation 
parameters \cite{carlos}: 
\be \label{eq:data}
\bad
\tan^2 \theta_{12} =  0.29 \ldots 0.82 ~,~ & 
\sin^2 \theta_{13} <   0.05  ~,~ & 
\tan^2 \theta_{23} =  0.45 \ldots 2.3   ~, \ea \ee
\be \nonumber 
\baz  
\dms \simeq (5.4 \, (14) \ldots 10 \, (19)) 
\times 10^{-5} \, \rm eV^2 ~,~ & 
\dma \simeq (1.5 \ldots 3.9) \times 10^{-3} \, \rm eV^2 ~, 
\ea
\ee
where the 90 \% C.L.\ ranges for the respective quantities are 
given. For \dms{} two upper and lower limits are given, 
corresponding to the so--called LMA--I and LMA--II solutions \cite{LMAI}. 
The best--fit points are located in the LMA--I parameter space 
and are \cite{carlos} 
$\tan^2\theta_{12} = 0.45$, $\dms = 7.1 \times 10^{-5}$ eV$^2$. 
For the atmospheric sector one finds the best--fit points 
$\tan^2\theta_{23} = 1 $ and $\dma = 2.6 \times 10^{-3}$ 
eV$^2$ \cite{carlos}. 
At the moment no information at all about the $CP$ phases or relative $CP$ 
parities exists. \\ 

\noindent Regarding the total neutrino mass scale, only upper limits exist. 
Three different observables are at one's disposal, 
the effective Majorana mass \meff{} as 
measurable in neutrinoless double beta decay, the sum of neutrino 
masses $\Sigma$ as testable through cosmology and 
the mass parameter $m_{\nu_e}$ 
as testable in direct kinematical experiments. Their 
definitions and current limits read  
\bea
\meff \equiv |\sum U_{ei}^2 \, m_i| \ls 1 \eV \, \, \cite{0vbb}~, \\[0.3cm]
\Sigma \equiv \sum |m_i| < 1.01 \eV \, \, \cite{Sigma}~, \\[0.3cm]
m_{\nu_e} \equiv  \sqrt{\sum |U_{ei}^2 \, m_i^2|} < 2.2 \eV \, \, 
\cite{trit}~.
\eea
Regarding \meff, a factor of $\sim 3$ for the uncertainty in the nuclear 
matrix element calculations was included. The Heidelberg--Moscow 
collaboration gives --- using the results of one specific 
calculation for the nuclear matrix elements --- 
a limit of $\meff < 0.35 $ eV at $90 \%$ C.L.~\cite{0vbb}.

\subsection{\label{sec:theory}Theory}
This letter is supposed to analyze the impact of a traceless $m_\nu$. 
There exists a very simple and phenomenologically highly 
interesting explanation for this possibility \cite{Tr1}. 
The neutrino mass matrix is given by the see--saw mechanism 
\cite{seesaw} in general as 
\be 
\label{eq:seesaw}
m_\nu =  M_L - m_D \, M_R^{-1} \, m_D^T~, 
\ee
where $m_D$ is a Dirac mass matrix and $M_R$ ($M_L$) a  
right--handed (left--handed) Majorana mass matrix. 
In $SO(10)$ models, 
choosing Higgs fields in ${\bf 10}$ and  ${\bf \overline{126}}$ and the 
$B - L$ breaking being performed by a ${\bf 126}$ Higgs, 
one can write (see e.g.\ \cite{SO10}): 
\be \label{eq:so10}
\baz 
M_L = Y_{126} \, v_L & M_R = Y_{126} \, v_R \\[0.3cm]
m_{\rm down} = Y_{10} \, v_{10}^{\rm down} 
+ Y_{126} \, v_{126}^{\rm down} & 
m_{\rm lep} = Y_{10} \, v_{10}^{\rm down} 
-3 \, Y_{126} \, v_{126}^{\rm down}\\[0.3cm]
m_{\rm up} = Y_{10} \, v_{10}^{\rm up} 
+ Y_{126} \, v_{126}^{\rm up} & 
m_{D} = Y_{10} \, v_{10}^{\rm up} 
-3 \, Y_{126} \, v_{126}^{\rm up}~,
\ea
\ee
where $m_{\rm down \, (lep)}$ are the down quark (charged lepton), 
$m_{\rm up \, (D)}$ the up quark (Dirac) mass matrices,  
$Y_{10}$ and $Y_{126}$ are the Yukawa coupling matrices 
and $v_{10, 126}^{\rm down\, (up)}$ are the vevs of the 
corresponding Higgs fields. 
The vevs corresponding to the Majorana mass matrices 
are denoted $v_L$ and $v_R$. 
From (\ref{eq:so10}) one finds 
$4 \, Y_{126} = (m_{\rm down} - m_{\rm lep})/v_{10}^{\rm down} $.
Suppose now that the first term in the see--saw formula (\ref{eq:seesaw})
dominates. The mass matrix reads in this case:  
\be \label{eq:typeII}
m_\nu = Y_{126} \, v_L = (m_{\rm down} - m_{\rm lep}) 
\frac{v_L}{4 \, v_{10}^{\rm down}}~.
\ee

\noindent Suppose now that $m_{\rm down}$ and $ m_{\rm lep}$ 
are hierarchical, 
i.e., they contain small off--diagonal entries and 
the diagonal entries correspond roughly to the down quark and charged lepton 
masses, respectively. Then, the 23 sector of $m_\nu$ is diagonalized by 
\be
\tan 2 \theta_{23} \propto \frac{1}{m_b - m_\tau}~.
\ee
This mixing becomes maximal when 
$b-\tau$ unification takes place, i.e., $m_b = m_\tau$. This simple and 
appealing argument was first given in \cite{btau1}. In \cite{btau2} the 
idea was generalized to the full 3 flavor case and shown to be fully 
consistent with existing neutrino data.\\ 

\noindent 
Here we wish to emphasize that due to the same fact, $b-\tau$ unification, 
the trace of $m_\nu$ is proportional to $m_b - m_\tau$ and therefore, 
to a good precision, the trace vanishes \cite{Tr1}. 
We can quantify the smallness of the trace as 
\be
{\rm Tr} \, m_\nu \simeq \frac{(m_s - m_\mu) \, v_L}{4 \, v_{10}^{\rm down}} 
\simeq 0.025 \, \left( \frac{v_L}{\rm eV}\right)
\, \left( \frac{\rm GeV}{v_{10}^{\rm down}}\right) \, \eV ~.
\ee
Here, $m_s \simeq 0.2 $ GeV and $m_\mu \simeq 0.1 $ GeV are the 
masses of the strange quark and muon, respectively. 
For the typical values of \cite{Mohbook} 
$v_L \ls 0.1$ eV and $v_{10}^{\rm down} \gs 1$ GeV we can expect 
the trace to be less than $10^{-3}$ eV\@. 
We shall take the fact as the starting point of our purely 
phenomenological analysis. Note that most of 
our results should be 
a specific case of a more detailed, but model--dependent analysis 
as performed in \cite{btau2}. They can serve as a simple insight 
of the physics involved and results obtained.

\section{\label{sec:CPcons}The $CP$ conserving case}
We shall investigate now the consequences of the requirement 
${\rm Tr} \, m_\nu = 0$ on the mass states in the $CP$ conserving case. 

\subsection{\label{sec:consNH}Normal hierarchy}
Allowing for arbitrary relative signs of the mass 
states $m_i$ with the convention $|m_3| > |m_2| > |m_1|$,  
the condition $m_1 + m_2 + m_3 = 0$ together with the experimental 
constraints of $\Delta m^2_{32} = \dma =  2.6 \times 10^{-3} \, \rm eV^2$ 
and $\Delta m^2_{21} = \dms = 7.1 \times 10^{-5} \, \rm eV^2$ is 
solved by 
\be
m_1 = 0.0290 \eV \simeq m_2 = 0.0302 \eV 
\mbox{ and } m_3 = -0.0593 \eV \simeq -2 \, m_2  ~.
\ee
The numbers of course coincide with the ones presented in \cite{Tr2}. 
The characteristic relation $|m_3| \simeq 2 \, |m_2| \simeq 2 \, |m_1|$ 
holds as long as ${\rm Tr} \, m_\nu \ls 10^{-3}$ eV\@. The mass 
spectrum corresponds to a ``partially degenerate'' scheme.\\

\noindent The different relative signs of the mass states correspond to the 
\mmp configuration, for which $\alpha = 2\beta = \pi$. 
The effective mass \meff{} reads for these values and for 
$|m_3| \simeq 2 \, |m_2| \simeq 2 \, |m_1|$  
\be \label{eq:meffNH0}
\meff \simeq \frac{|m_3|}{4} \, 
\left(3 \, \cos 2 \theta_{13} - 1 \right)~,  
\ee
which is independent on $\tan^2 \theta_{12}$. Varying $\theta_{13}$ 
leads to values of 
$0.025 \eV \! \ls \meff \ls 0.030$ eV, thus predicting a very narrow range 
within the reach of running and future experiments \cite{ovbbrev}.

\noindent Direct kinematical measurements will have to 
measure 
\be
m_{\nu_e} \simeq \frac{|m_3|}{2} \, \sqrt{1 + 4 \, \sin^2 \theta_{13}} 
\simeq (0.030 \ldots 0.032) \eV~,
\ee
which is one order of magnitude 
below the limit of the future KATRIN experiment \cite{KATRIN}.

\noindent The sum of the absolute values of the neutrino masses 
is $\Sigma  \simeq 0.12$ eV. 
As shown in \cite{han}, this is the lowest 
value (at 95 \% C.L.) measurable by 
combining data from the PLANCK satellite experiment and the 
Sloan Digital Sky Survey. 
Galaxy surveys one order of magnitude larger could reduce this limit by 
a factor of two \cite{han} and thus test the prediction.\\

\noindent We turn now to a simple form of the mass matrix that 
accomplishes the requirement of being traceless. 
We concentrate on mass matrices with three parameters, sizable 
$U_{e3}$ and no zero entries. 
For hierarchical neutrinos one might expect a quasi--degenerated and 
dominant 23 block of the mass matrix. Thus, one is lead to propose
\be \label{eq:mnuNH1}
m_\nu = 
\left( 
\bad 
-a & \epsilon_1 & \epsilon_2 \\[0.3cm]
\cdot & a/2 & 3a/2 \\[0.3cm]
\cdot & \cdot & a/2 
\ea
\right)~,
\ee
where $\epsilon_{1,2} \ll a$. 
Note that with $\epsilon_i = 0$ 
the mixing angles are $\theta_{23} = \pi/4$, $\theta_{13} = 0$ 
and $\tan \theta_{12} = 1/\sqrt{2}$, which is a widely discussed 
scheme \cite{bitri}. 
We find with the mass matrix (\ref{eq:mnuNH1}) that the mass states 
are 
\be 
\bad \D 
m_3 \simeq 2 \, a~, & \D m_2 \simeq -a - 
\frac{\epsilon_1 - \epsilon_2}{\sqrt{2}}~, 
& \D m_2 \simeq -a + \frac{\epsilon_1 - \epsilon_2}{\sqrt{2}}~,
\ea 
\ee
and the observables are given by 
\be \label{eq:resNH1}
\baz \D \D
\dma \simeq 3 \, a^2 ~,& \D \dms \simeq 2 \, \sqrt{2} \, a \, 
(\epsilon_1 - \epsilon_2)~,
 \\[0.3cm] \D 
\tan 2 \theta_{12} \simeq 6 \, \sqrt{2} \, a \, 
\frac{\epsilon_1 - \epsilon_2}{(\epsilon_1 + \epsilon_2)^2}~, & \D 
\sin \theta_{13} \simeq \frac{1}{3 \, \sqrt{2} \, a} \, 
(\epsilon_1 + \epsilon_2 )~,
\ea 
\ee
together with maximal atmospheric mixing. 
For $\epsilon_1 = \epsilon_2 \neq 0$ the solar mixing angle 
vanishes. 
Comparing the last two equations with the data from Eq.\ (\ref{eq:data}), 
one finds that $a^2 \simeq 10^{-3} \eV^2$, $\epsilon_1 - \epsilon_2 \simeq 
10^{-3} \eV$ and $\epsilon_1 + \epsilon_2 \simeq 10^{-2} \eV$ in 
order to reproduce the observations. It is seen that 
$|U_{e3}|$ should be sizable; we can express this element  
in terms of the other observables as 
\be \label{eq:corNH}
U_{e3}^2 \simeq \frac{1}{4} \, \frac{\dms}{\dma} \, 
\frac{1 - \tan^2 \theta_{12}}{\tan \theta_{12}}~, 
\ee
which becomes smaller, the larger the solar neutrino mixing angle 
$\theta_{12}$ becomes. Inserting the data from Eq.\ (\ref{eq:data}) 
in the right--hand side of the equation, the range 
of $U_{e3}^2$ is found for the LMA--I (LMA--II) case to lie 
between 0.0007 (0.002) and 0.0022 (0.04) in accordance with its current 
limit. The best--fit point predicts 
$U_{e3}^2 \simeq 0.0056$. The effective mass is given by 
\be \label{eq:meffNH}
\meff = a \simeq \sqrt{\dma/3} \sim 0.03 \eV~, 
\ee
where we inserted the best--fit value of \dma. 
The allowed range of \meff{}  
lies between 0.022 eV and 0.036 eV, with a  
best--fit prediction of 0.029 eV\@.  
Both observables should thus be measurable 
with the next round of experiments. An alternative 
formulation of the correlation of observables reads 
\be \label{eq:corNH2}
U_{e3}^2 \simeq \frac{\dms}{12 \, \meff^2} \, 
\frac{1 - \tan^2 \theta_{12}}{\tan \theta_{12}}~,
\ee
which could be used as a further check if both \meff{} and $U_{e3}^2$ were 
measured.\\

\noindent 
The question arises if the results are stable against radiative corrections. 
As known, the 12 sector is unstable for quasi--degenerate 
neutrinos with equal relative $CP$ parity \cite{rad}, which is what 
happens here. 
The effect of radiative corrections can be estimated by multiplying  
the $\alpha \beta$ element of $m_\nu$ with a term 
$(1 + \delta_\alpha) \, (1 + \delta_\beta)$, where 
\be
\delta_\alpha = c \, \frac{m_\alpha^2}{16 \, \pi^2 \, v^2} 
\, \ln \frac{M_X}{m_Z}~.
\ee   
Here $m_\alpha$ is the mass of the corresponding charged lepton, 
$M_X \simeq 10^{16}$ GeV and $c = -(1 + \tan^2 \beta)$ (3/2) in case of the 
MSSM (SM). We checked numerically that  
for the SM there is no significant change of $\theta_{12}$ and \dms{} 
but for the MSSM and $\tan \beta \gs 20$ the 
results become unstable.  
Also, the relation between $|U_{e3}|$ and the other 
observables remains its validity for the SM and for the MSSM 
with $\tan \beta \ls 20$.\\

\noindent One can relax the traceless condition a bit by adding a term 
proportional to $\mathbbm{1} \times \xi/3 $ to $m_\nu$, 
where $\xi = $ Tr $m_\nu$. The mixing angles are of course 
unaffected by this term but the masses are changed by 
$m_i \rightarrow m_i + \xi/3$. Thus, the new mass squared 
differences read 
\be
\dms \simeq 2 \, \sqrt{2} \, (\epsilon_1 - \epsilon_2) \, (a - \xi/3) 
\mbox{ and } \dma \simeq 3 \, a \, \left(a + \frac{2}{3} \, \xi\right)~.
\ee
The correlation of the observables $U_{e3}$ and \meff{} also 
changes, it is now given by  
\bea \D
U_{e3}^2 \simeq \frac{1}{4} \, \dms \, 
\frac{1 - \tan^2 \theta_{12}}{\tan \theta_{12}}
\frac{1}{\dma + \xi (\xi - \sqrt{\xi^2 + 3 \, \dma})}~, \\[0.4cm] \D 
\meff = a - \xi/3 \simeq \frac{1}{3} \, 
\left( \sqrt{\xi^2 + 3 \, \dma} - 2 \, \xi
\right)~.
\eea
For $\xi = 0$ the previous two equations reproduce 
(\ref{eq:resNH1},\ref{eq:corNH},\ref{eq:meffNH}). 
The formula for the correlations of the 
observables, Eq.\ (\ref{eq:corNH2}), holds also in the case of $\xi \neq 0$. 
As long as $\xi$ does not exceed $10^{-3}$ eV, 
the corrections due to Tr $m_\nu = \xi \neq 0$ increase (reduce) the 
predictions for $U_{e3}^2$ (\meff) by $\sim \xi/\sqrt{\dma} \simeq 
3 \, \% $. 


\subsection{\label{sec:consIH}Inverted hierarchy}
Allowing for arbitrary relative signs of the mass 
states $m_i$ together with the convention $|m_2| > |m_1| > |m_3|$, 
the condition $m_1 + m_2 + m_3 = 0$ with the experimental 
constraints of $\Delta m^2_{13} = \dma = 2.6 \times 10^{-3} \, \rm eV^2$ 
and $\Delta m^2_{21} = \dms = 7.1 \times 10^{-5} \, \rm eV^2$ is 
solved by 
\be
m_2 = 0.0517 \eV \simeq -m_1 = 0.0510 \eV 
\mbox{ and } m_3 = -0.0007 \eV~.
\ee
The 
characteristic relation $|m_2| \simeq |m_1| \gg |m_3|$  
holds as long as ${\rm Tr} \, m_\nu \ls 10^{-2}$ eV\@.\\

\noindent The signs of the mass states correspond to the 
\pmm configuration for which $\alpha = \beta = \pi/2$. The 
effective mass then reads for 
$|m_2| \simeq |m_1| \gg |m_3|$
\be \label{eq:meffIH0}
\meff \simeq |m_2| \, \cos2 \theta_{12} \, \cos^2 \theta_{13}  
\simeq |m_2| \, \frac{1 - \tan^2 \theta_{12}}{1 + \tan^2 \theta_{12}}~,
\ee
which lies between 0.005 eV and 0.028 eV, thus predicting a range 
with the upper (lower) limit 
within (outside) 
the reach of running and future experiments \cite{ovbbrev}. 
The 
lower limit is however reachable by the 10t version of the 
GENIUS \cite{GENIUS} project. 
The best--fit prediction is $\meff \simeq 0.020$ eV\@. 
In contrast to 
the normal mass ordering, \meff{} has a crucial dependence 
on $\tan^2 \theta_{12}$ and thus a rather large allowed range.

\noindent The mass measured in direct kinematical experiments 
is $m_{\nu_e} \simeq |m_2| \simeq 0.05$ eV, which is larger than 
the corresponding quantity in the normal hierarchy but still almost 
one order of magnitude below the limit of the future KATRIN experiment.

\noindent The sum of the absolute values of the neutrino masses 
is $\Sigma  \simeq 0.10$ eV, lower than the corresponding quantity 
in the normal hierarchy and thus still requiring larger galaxy surveys, 
as shown in \cite{han}.\\

\noindent We present again a simple 3 parameter 
mass matrix with the traceless feature, no zero entries and 
non--vanishing $U_{e3}$. One is naturally lead to propose 
\be \label{eq:mnuIH1}
m_\nu = 
\left( 
\bad 
-a & b  & -b  \\[0.3cm]
\cdot & a/2 - \eta & -a/2 \\[0.3cm]
\cdot & \cdot & a/2 + \eta
\ea
\right)~,
\ee
where $b > a > \eta$. 
We find with the mass matrix (\ref{eq:mnuNH1}) that the mass states 
for $b^2 \gg \eta^2$ are 
\be
m_{2,1} \simeq \pm \frac{8 \, b^4 + a^2 \, 
\left(4 \, b^2 + \left(1 \pm \sqrt{1 + 2 \, b^2/a^2} \right) \, \eta^2 \right)}
{4 \, b^2 \sqrt{a^2 + 2 \, b^2}} \mbox{ and } 
m_3 \simeq \frac{-a}{2 \, b^2} \, \eta^2~.
\ee
The observables are found to be 
\be \label{eq:resIH1}
\baz \D 
\tan 2 \theta_{12} \simeq \sqrt{2} \, \frac{b}{a}~, & \D 
\sin \theta_{13} \simeq \frac{\eta}{\sqrt{2} \, b}~,
\\[0.3cm] \D 
\dma \simeq a^2 + 2 \, b^2 ~,& 
\D \dms \simeq \frac{a}{b^2} \, \sqrt{a^2 + 2 \, b^2} \, \eta^2~.
\ea 
\ee
Again, the observed values 
of the quantities are easy to reproduce with, in this case, e.g., 
$b > a > \eta \sim 0.01$ eV\@.
 
\noindent There is again a simple correlation of the 
observables, namely 
\be \label{eq:corIH1}
U_{e3}^2 \simeq \frac{1}{2} \, \frac{\dms}{\dma} \, 
\frac{1 + \tan^2 \theta_{12}}{1 - \tan^2 \theta_{12}}~. 
\ee
Putting again the data from Eq.\ (\ref{eq:data}) in the right--hand side of 
this equation leads to $U_{e3}^2 \gs 0.013$ with a best--fit 
prediction of $U_{e3}^2 \simeq 0.036$. For large values of \dms, i.e., 
in the less favored 
LMA--II solution, which corresponds to 
$\dms \gs 10^{-4} \eV^2$, the value of $U_{e3}$ is above its current 
experimental limit. Comparing the expressions for $U_{e3}$ in the 
normal (Eq.\ (\ref{eq:corNH})) and inverted (Eq.\ (\ref{eq:corIH1}))
ordering leads to the observation that for the first case the value is 
smaller by a factor of 
$\simeq 1/2 \, (1 - \tan^2 \theta_{12})^2 
/((1 + \tan^2 \theta_{12}) \tan \theta_{12}) \ls 0.35$.

\noindent The effective mass is given by 
\be \label{eq:IHmeff}
\meff = a 
\simeq \frac{\dms}{2 \, \sqrt{\dma} \, U_{e3}^2} 
= \sqrt{\dma} \, \frac{1 - \tan^2 \theta_{12}}{1 + \tan^2 \theta_{12}} 
\sim 0.02 \eV~. 
\ee
Comparing this result with our prediction for \meff{} in the 
normal mass ordering, Eq.\ (\ref{eq:resNH1}), one finds that 
the inverted mass hierarchy predicts an effective mass smaller than a factor 
$\sqrt{3} \tan 2 \theta_{12} \gs 4$. This is larger than the typical 
uncertainty of the nuclear matrix elements that usually tends to spoil 
extraction of information from \onbb$\!\!$.
 
\noindent If both, \meff{} and $U_{e3}$ are measured, 
one can further check the mass matrix by the relation 
\be \label{eq:corIH2}
U_{e3}^2 \simeq \frac{\dms}{2 \, \meff{} \, \sqrt{\dma} }~.
\ee
We checked numerically 
that the results are stable under radiative corrections in the SM and 
in the MSSM for $\tan \beta \ls 50$.\\

\noindent One can again relax the traceless condition 
through a contribution $\mathbbm{1} \times \xi/3 $ to $m_\nu$, 
where $\xi = $ Tr $m_\nu$. The new mass squared 
differences are 
\be
\dms \simeq \frac{a \, \eta^2 + \frac{4}{3} \, b^2 \, \xi}{b^2} \, 
\sqrt{a^2 + 2 \, b^2} 
\mbox{ and } 
\dma \simeq a^2 + 2 \, b^2 - \frac{2}{3} \, \sqrt{a^2 + 2 \, b^2} \, \xi~.
\ee
Again, the correlation of the observables $U_{e3}$ and \meff{} 
changes, one finds 
\bea \D 
U_{e3}^2 \simeq \frac{9}{2} \, 
\frac{1 + \tan^2 \theta_{12}}{1 - \tan^2 \theta_{12}} 
\frac{\dms - \frac{4}{9} \, \xi (\xi + \sqrt{9 \, \dma + \xi^2})}
{(\xi + \sqrt{9 \, \dma + \xi^2})^2} \\[0.4cm] \D 
\meff = a - \xi/3 \simeq 
\frac{1}{3 \, (1 + \tan^2 \theta_{12})} \, \left( 
(1 - \tan^2 \theta_{12}) \, \sqrt{\xi^2 + 9 \, \dma} 
- 2 \, \tan^2 \theta_{12}\right) 
~.
\eea
For $\xi = 0$ the results for exact zero trace given above are 
re--obtained.\\

\noindent Interestingly, the same mass matrix, Eq.\ (\ref{eq:mnuIH1}), 
has been found in \cite{detmot2}. 
In this work a local horizontal $SU(2)$ symmetry 
has been applied to the charged leptons. A consequence was 
a vanishing determinant of $m_\nu$ and an inverted hierarchy 
for the neutrino masses (i.e.\ $m_3 = 0$) 
with opposite signs for $m_2$ and $m_1$. In this case, both the trace and 
the determinant of $m_\nu$ are vanishing, which explains that our results 
are identical to the ones in \cite{detmot2}.\\

\noindent To put this Section in a nutshell, the requirement of a vanishing 
trace of $m_\nu$ leads in the $CP$ conserving case to 
values of \meff, larger in the NH than in the IH. 
Due to the dependence on $\tan \theta_{12}$, \meff{} in case of IH 
has a broad range. 
Simple mass matrices were proposed which reproduce the values found by 
the traceless condition and in addition predict larger $U_{e3}$ in the 
IH case. 
Relaxing the traceless condition does not significantly
change the predicted values 
as long as the trace stays below the expected $10^{-3}$ eV.

\section{\label{sec:CPviol}The $CP$ violating case}
Now we shall investigate the more realistic case of 
$CP$ violation and the consequences of the traceless $m_\nu$ 
condition. Within the parametrization (\ref{eq:Upara}) one finds 
--- using Eq.\ (\ref{eq:mnu}) --- for the trace of $m_\nu$ that 
\be \label{eq:compl}
{\rm Tr \,} m_\nu \simeq m_1 + m_2 \, e^{2 i \alpha} \, + 
m_3 \, e^{2 i (\beta + \delta)}~. 
\ee
Terms of order $\sin^2 \theta_{13}$ were neglected, which can be shown to be 
a justified approximation. 
The condition of zero trace holds for the real and 
imaginary part of Tr $m_\nu$, i.e., 
\bea \label{eq:compl12}
m_1 + m_2 \, \cos 2 \alpha + m_3 \, \cos 2 (\beta + \delta) = 0 \\[0.3cm]
 m_2 \, \sin 2 \alpha + m_3 \, \sin 2 (\beta + \delta) = 0~.
\eea
The minimal values of $m_1$ or $m_3$ that fulfill the condition 
(\ref{eq:compl12}) are the ones that correspond to the $CP$ 
conserving case discussed in the previous Section.  
As a check, one can convince oneself that for $\delta = 0$ and 
$m_1 = m_2 = m_3/2$ the solution of the two 
equations in (\ref{eq:compl12}) is given by $\alpha = 2 \beta = \pi$ 
while for $\delta = 0$ and $m_1 = m_2 \gg m_3 \simeq 0$ one finds that 
$\alpha = \pi/2$, which is in accordance with the previous Section. 
This means that in case of the normal hierarchy and 
the LMA I (LMA II) solution a lower limit 
on the neutrino mass of 0.019 (0.021) eV can be set, which is obtained by 
inserting the lowest allowed \dma{} and the largest \dms. 
In case of inverted hierarchy, one finds that 
$|m_3| \ge 0.0013$ (0.0024) eV for the LMA I (LMA II) solution. 

\noindent Due to the zero trace condition one can write 
\be \label{eq:m123}
m_1^2 = m_2^2 + m_3^2 + 2 \, m_2 \, m_3 \, \cos \phi~  
\mbox{, where }~ \phi = 2(\alpha - \beta - \delta)~.
\ee 
Interestingly, this implies that in the expressions for 
$\Sigma$ and $m_{\nu_e}$ the phases appear. In particular, 
\be
m_{\nu_e}^2 = \frac{1}{1 + \tan^2 \theta_{12}} \, 
\left(
m_3^2 + m_2^2 \, (1 + \tan^2 \theta_{12}) + 2 \, m_2 \, m_3 \, 
\cos \phi 
\right) + m_3^2 \, \sin^2 \theta_{13}~.
\ee
For quasi--degenerate neutrinos, i.e., $m_3 \simeq m_2 \simeq m_1 
\equiv m_0$,  one finds from Eq.\ (\ref{eq:m123}) that 
$\cos \phi = -1/2$ or 
equivalently $\alpha - \beta - \delta \simeq \pm \pi/3 \pm n \pi$. Thus, 
quasi--degenerate neutrinos and the zero trace condition demand 
non--trivial correlations between the $CP$ phases.\\

\noindent Applying the condition Tr $m_\nu = 0$ to Eq.\ (\ref{eq:m123}) and 
inserting it in the expression for \meff{} one finds 
\be \D 
\meff \simeq \frac{1}{1 + \tan^2 \theta_{12}} \, 
\sqrt{m_3^2 + m_2 \, (1 - \tan^2 \theta_{12}) \, 
\left( 
m_2 \, (1 - \tan^2 \theta_{12}) + 2 \, m_3 \, \cos \phi
\right)}~,
\ee
where we neglected $\sin^2 \theta_{13}$. 
Courtesy of the zero trace condition, \meff{} depends 
effectively only on one phase.  
The $CP$ conserving cases in the previous Section should come as 
special cases of the last formula. Indeed, 
for $\delta = 0$,  
$m_2 = m_3/2$ and $\alpha = 2 \beta = \pi$ one recovers 
Eq.\ (\ref{eq:meffNH0}) and for $\delta = 0$ and $m_3 = 0$ one re--obtains 
Eq.\ (\ref{eq:meffIH0}).
For quasi--degenerate neutrinos $m_0 \equiv m_2 \simeq m_3$ 
the above formula simplifies. 
Then, since $\cos \phi \simeq -1/2$:  
\be
\meff \simeq m_0 \, 
\frac{\sqrt{1 + \tan^2 \theta_{12} \, (\tan^2 \theta_{12} - 1)}}
{1 + \tan^2 \theta_{12}}~,
\ee
which can be used to set an upper limit on $m_0$. 
For $\meff \ls 1$ eV we have $m_0 \ls 1.96$ eV with a best--fit limit 
of 1.67 eV. 
Therefore, the zero trace condition implies a limit stronger than the 
one stemming from direct kinematical experiments. Using the less 
conservative limit given by the Heidelberg--Moscow collaboration, 
the above limits are reduced by a factor of roughly 2.9 and 
the limits come nearer to the ones from cosmological observations. 
To be precise, for $\meff < 0.35$ eV one finds $m_0 \ls 0.69$ eV and, 
for the best--fit value, $m_0 \ls 0.58$ eV. 
The values are testable by the KATRIN experiment. Thus, together with 
the lower limit (about 0.02 eV for NH and 0.002 eV for IH) 
from the beginning of this Section, a neutrino mass window is defined.\\
 

\noindent One can 
compare the predictions for $\meff$ in case of zero trace 
with the ones in case of zero determinant \cite{det}. 
This corresponds for NH (IH) to zero $m_1$ ($m_3$), which results 
in particular simple forms of \meff, see \cite{det} for details. 
Regarding the inverted hierarchy, we already commented 
that in case of an opposite relative sign of the 
two quasi--degenerate neutrinos and a very small $m_3$ 
both the trace and the determinant vanish and the situation is identical.  
We use the data from Eq.\ (\ref{eq:data}) for our predictions. 
For the normal mass ordering strong cancellations are 
possible \cite{ichNPB,cancel}, and \meff{} is 
in general predicted to be below 0.01 eV. 
In case of the inverted mass ordering, 
\meff{} lies between 0.004 and 0.034 eV, independent on 
$\sin^2 \theta_{13}$. 
Unlike the zero trace case, the zero determinant conditions allows no 
statements about the phases, at least not before the limit on \meff{} 
is significantly improved.\\

\noindent We also 
performed a numerical analysis of the zero trace condition. 
For this exercise the mass squared differences and 
solar neutrino mixing angle were fixed to 
their best--fit points and the 
smallest neutrino mass and the phases $\alpha$ as well as 
$\beta - \delta$ were 
varied within their allowed range. 
The results in the form of scatter plots for the normal hierarchy is shown 
in Fig.\ \ref{fig:NH} and for the inverted scheme in Fig.\ \ref{fig:IH}. 
One recognizes for example in Figs.\ (\ref{fig:NH},\ref{fig:IH})c 
the correlation of $\Sigma$ with 
$\alpha - \beta - \delta$ as implied by Eq.\ (\ref{eq:m123}).  
For the inverted hierarchy, the spread of the phases 
is rather different from the case of normal hierarchy. 
This can be understood from the fact 
that for small $m_3$ the dependence on $\beta - \delta$ 
practically vanishes.

\section{\label{sec:concl}Summary and Conclusions}
\begin{figure}
\begin{center}
\hspace{-3cm}\vspace{-2cm}
\epsfig{file=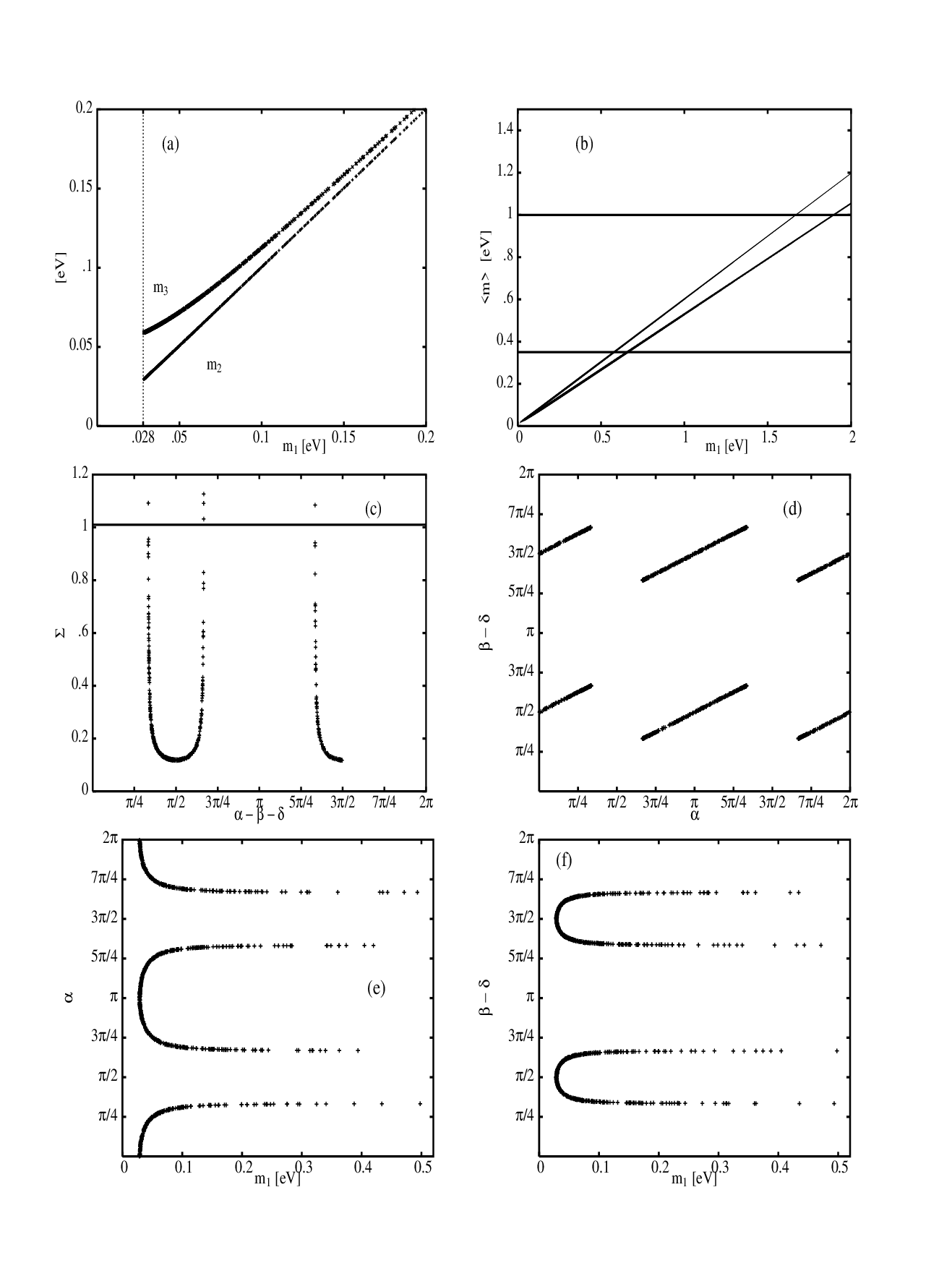,width=19cm,height=22cm}
\caption{\label{fig:NH}Scatter plot of different parameters in the 
normal mass ordering obtained by 
varying the smallest mass state $m_1$ and the phases. 
The oscillation parameters were set to their best--fit values. 
Shown are 
(a) $m_1$ against the two other two masses, 
(b) $m_1$ against the minimal and maximal value of \meff{} (given 
by varying $\theta_{13}$), 
(c) $\alpha - \beta$ against $\Sigma$, 
(d) $\alpha$ against $\beta$, 
(e) $m_1$ against $\alpha$ and 
(f) $m_1$ against $\beta$.}
\end{center}
\end{figure}

\begin{figure}
\begin{center}
\hspace{-3cm}
\epsfig{file=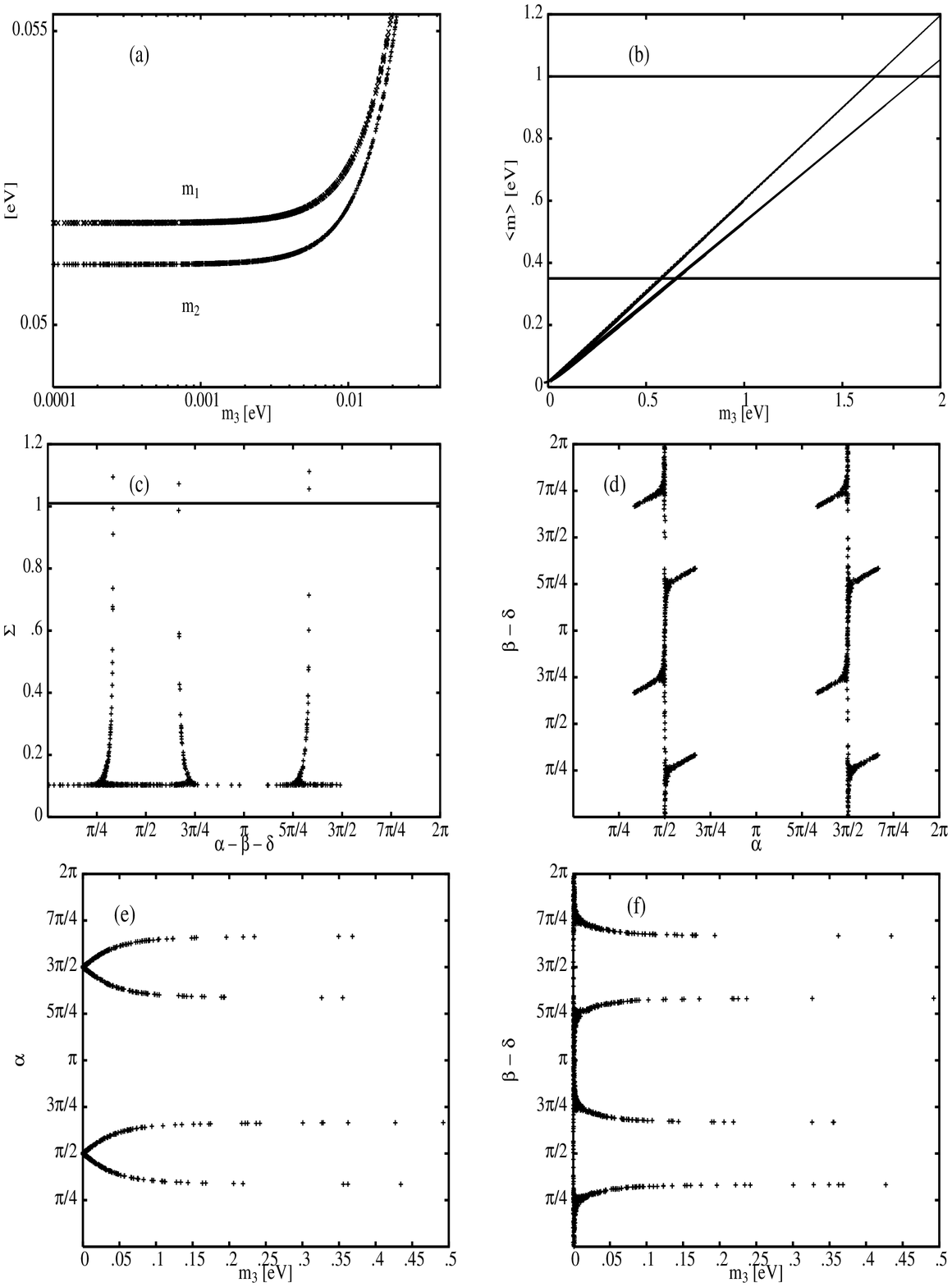,width=19cm,height=22cm}
\caption{\label{fig:IH}Same as previuos figure for the inverted mass 
ordering.}
\end{center}
\end{figure}

The condition of a zero trace of the neutrino mass matrix $m_\nu$ 
was reanalyzed in case of $CP$ conservation and violation for both 
possible mass orderings. 
The motivation for this purely phenomenological analysis 
was given by exact $b-\tau$ unification in connection to 
the non--canonical type II see--saw 
mechanism in $SO(10)$ models. This situation has 
recently gathered renewed attention because of its ability to 
produce large atmospheric neutrino mixing in a simple way.  
In case of $CP$ conservation, the values of the neutrino masses 
and their relative $CP$ parities are fixed and allow to give 
simple expressions for the effective mass as measurable in \onbb$\!\!$. 
The masses are given by $m_1 \simeq m_2 \simeq -m_3/2 \simeq 0.03$ eV for 
the normal mass ordering and $\sqrt{\dma} \simeq 
m_2 \simeq - m_1 \gg- m_3$ for the 
inverted mass ordering. 
In case of the normal hierarchy, \meff{} does not depend on the 
solar neutrino mixing angle and is predicted to be 
around 0.03 eV. In case of inverted hierarchy, \meff{} depends rather 
strongly on the solar neutrino mixing angle and its range is between 
0.005 eV and 0.03 eV; the best--fit prediction is 0.02 eV. 
The presence of $CP$ violation and therefore non--trivial values of the 
Majorana phases allows for larger values of the masses. In case of 
quasi--degenerate neutrinos a peculiar relation between the 
phases exists: $\alpha - \beta - \delta = \pm \pi/3 \pm n \pi$. 
The minimal 
values of the masses correspond to the $CP$ conserving case and are 
in case of the normal (inverted) hierarchy roughly 0.02 (0.002) eV. 
The upper limit comes from non--observation of neutrinoless 
double beta decay and is for $\meff < 0.35$ eV roughly 0.7 eV. 
Correlations of various parameters are possible, some of which 
are shown in Figs.\ \ref{fig:NH} and \ref{fig:IH}.

\vspace{0.5cm}
\begin{center}
{\bf Acknowledgments}
\end{center}
I thank R.~Mohapatra and S.T.~Petcov 
for helpful discussions and careful reading of the 
manuscript. 
Part of this work was performed at the 
Baryogenesis Workshop at the MCTP in Ann Arbor, Michigan.  
I wish to thank the organizers for the stimulating 
atmosphere they created and for financial support. 
The hospitality of the DESY theory group, where other parts of the work were 
performed, is gratefully acknowledged. 
This work was supported by the EC network HPRN-CT-2000-00152.

\newpage


\pagestyle{empty}


\begin{thebibliography}{99} 
\bibitem{ichCP}S.~Pascoli, S.~T.~Petcov and L.~Wolfenstein, 
Phys.\ Lett.\ B {\bf 524} (2002) 319; 
W.~Rodejohann, hep-ph/0203214; 
a very pessimistic view is given in 
V.~Barger, {\it et al.}, 
Phys.\ Lett.\ B {\bf 540} (2002) 247,   
a more optimistic one in 
S.~Pascoli, S.~T.~Petcov and W.~Rodejohann, 
Phys.\ Lett.\ B {\bf 549} (2002) 177. 
\bibitem{zeros}P.~H.~Frampton, S.~L.~Glashow and D.~Marfatia, 
Phys.\ Lett.\ B {\bf 536} (2002) 79; 
Z.~Z.~Xing, Phys.\ Lett.\ B {\bf 530}, 159 (2002); 
Phys.\ Lett.\ B {\bf 539}, 85 (2002); 
P.~H.~Frampton, M.~C.~Oh and T.~Yoshikawa, 
Phys.\ Rev.\ D {\bf 66} (2002) 033007; 
A.~Kageyama, {\it et al.}, Phys.\ Lett.\ B {\bf 538}, 96 (2002); 
B.~R.~Desai, D.~P.~Roy and A.~R.~Vaucher, Mod.\ Phys.\ Lett.\ A {\bf 18} 
(2003) 1355. 
\bibitem{det}G.~C.~Branco, {\it et al.}, 
Phys.\ Lett.\ B {\bf 562} (2003) 265.  
\bibitem{detmot1}See, e.g., 
W.~Grimus and L.~Lavoura, 
Phys.\ Rev.\ D {\bf 62} (2000) 093012; 
T.~Asaka, {\it et al.}, 
Phys.\ Rev.\ D {\bf 62} (2000) 123514; 
R.~Kuchimanchi and R.~N.~Mohapatra, 
Phys.\ Rev.\ D {\bf 66}, 051301 (2002); 
P.~H.~Frampton, S.~L.~Glashow and T.~Yanagida, 
Phys.\ Lett.\ B {\bf 548} (2002) 119; 
T.~Endoh, {\it et al.}, 
Phys.\ Rev.\ Lett.\  {\bf 89} (2002) 231601; 
M.~Raidal and A.~Strumia,
Phys.\ Lett.\ B {\bf 553} (2003) 72; 
B.~Dutta and R.~N.~Mohapatra, hep-ph/0305059. 
\bibitem{detmot2}R.~Kuchimanchi and R.~N.~Mohapatra, 
Phys.\ Lett.\ B {\bf 552} (2003) 198.
\bibitem{Tr1}D.~Black, {\it et al.}, 
Phys.\ Rev.\ D {\bf 62} (2000) 073015;
\bibitem{Tr2}X.~G.~He and A.~Zee, hep-ph/0302201.
\bibitem{modrad}A.~Zee,
Phys.\ Lett.\ B {\bf 93} (1980) 389
[Erratum-ibid.\ B {\bf 95} (1980) 461]; 
L.~Wolfenstein, 
Nucl.\ Phys.\ B {\bf 175} (1980) 93.
\bibitem{typeII}
R.~N.~Mohapatra and G.~Senjanovic, 
Phys.\ Rev.\ D {\bf 23}, 165 (1981); 
C.~Wetterich,
Nucl.\ Phys.\ B {\bf 187}, 343 (1981); 
J.~C.~Montero, C.~A.~de S. Pires and V.~Pleitez,
Phys.\ Lett.\ B {\bf 502} (2001) 167; 
R.~N.~Mohapatra, A.~Perez-Lorenzana and C.~A.~de Sousa Pires,
Phys.\ Lett.\ B {\bf 474}, 355 (2000); 
\bibitem{btau1}B.~Bajc, G.~Senjanovic and F.~Vissani, hep-ph/0110310; 
Phys.\ Rev.\ Lett.\  {\bf 90} (2003) 051802. 
\bibitem{btau2}H.~S.~Goh, R.~N.~Mohapatra and S.~P.~Ng, hep-ph/0303055.
\bibitem{PMNS}B. Pontecorvo, Zh.\ Eksp.\ Teor.\ Fiz.\ {\bf 33}, 549 (1957) 
and {\bf 34}, 247 (1958); 
Z. Maki, M. Nakagawa and S. Sakata, 
Prog.\ Theor.\ Phys.\ {\bf 28}, 870 (1962).
\bibitem{STP}S.~M.\ Bilenky {\it et al.}, Phys.\ Lett.\ B  {\bf 94}, 
495 (1980); 
M.~Doi, {\it et al.}, Phys.\ Lett.\ B {\bf 102}, 323 (1981); 
J.~Schechter and J.~W.~Valle, 
Phys.\ Rev.\ D {\bf 22}, 2227 (1980).
\bibitem{ichJPG}W.~Rodejohann, 
J.\ Phys.\ G {\bf 28} (2002) 1477.
\bibitem{mnu}M.~Frigerio and A.~Y.~Smirnov, 
Nucl.\ Phys.\ B {\bf 640} (2002) 233; 
Phys.\ Rev.\ D {\bf 67} (2003) 013007. 
\bibitem{ichPRD}W.~Rodejohann,
Phys.\ Rev.\ D {\bf 62} (2000) 013011. 
\bibitem{phaalt}L.~Wolfenstein, 
Phys.\ Lett.\ B {\bf 107} (1981) 77; 
S.~M.~Bilenky, N.~P.~Nedelcheva and S.~T.~Petcov, 
Nucl.\ Phys.\ B {\bf 247}, 61 (1984);
B.~Kayser, 
Phys.\ Rev.\ D {\bf 30}, 1023 (1984).
\bibitem{ichNPB}W.~Rodejohann, 
Nucl.\ Phys.\ B {\bf 597}, 110 (2001).
\bibitem{carlos}M.~C.~Gonzalez-Garcia and C.~Pena-Garay, hep-ph/0306001.
\bibitem{LMAI}
G.~L.~Fogli, E.~Lisi, A.~Marrone, D.~Montanino, A.~Palazzo and A.~M.~Rotunno,
Phys.\ Rev.\ D {\bf 67} (2003) 073002; 
A.~Bandyopadhyay, S.~Choubey, R.~Gandhi, S.~Goswami and D.~P.~Roy,
Phys.\ Lett.\ B {\bf 559} (2003) 121; 
J.~N.~Bahcall, M.~C.~Gonzalez-Garcia and C.~Pe{\~n}a-Garay,
JHEP {\bf 0302} (2003) 009; 
P.~C.~de Holanda and A.~Y.~Smirnov, 
JCAP {\bf 0302} (2003) 001.
\bibitem{0vbb}H.~V.~Klapdor-Kleingrothaus {\it et al.},
Eur.\ Phys.\ J.\ A {\bf 12} (2001) 147.
\bibitem{Sigma}S.~Hannestad, astro-ph/0303076.
\bibitem{trit}C.~Weinheimer, hep-ex/0306057.
\bibitem{seesaw}M. Gell--Mann, P. Ramond, 
and R. Slansky in {\it Supergravity},
p. 315, edited by F. Nieuwenhuizen 
and D. Friedman, North Holland,
Amsterdam, 1979;
T. Yanagida, Proc. of the 
{\it Workshop on Unified Theories and the Baryon
Number of the Universe}, edited by 
O. Sawada and A. Sugamoto, KEK, Japan 1979;
R.N. Mohapatra, G. Senjanovic, \Jo{\PRL}{44}{912}{1980}.
\bibitem{SO10}See, e.g., K.~S.~Babu and R.~N.~Mohapatra,
Phys.\ Rev.\ Lett.\  {\bf 70}, 2845 (1993); 
K.~Matsuda, T.~Fukuyama and H.~Nishiura,
Phys.\ Rev.\ D {\bf 61} (2000) 053001. 
\bibitem{Mohbook}See, e.g., 
R.~N.~Mohapatra, Unification and supersymmetry : 
The Frontiers of quark -- lepton physics, Springer-Verlag, 3rd edition 2003. 
\bibitem{ovbbrev}O. Cremonesi, Invited talk at the 
XXth Internat. Conf. on Neutrino Physics and Astrophysics (Neutrino 2002),
Munich, Germany, May 25-30, 2002, hep-ex/0210007.
\bibitem{KATRIN}A.~Osipowicz {\it et al.}  [KATRIN Collaboration], 
hep-ex/0109033.
\bibitem{han}S.~Hannestad, 
Phys.\ Rev.\ D {\bf 67} (2003) 085017.
\bibitem{bitri}L.~Wolfenstein,
Phys.\ Rev.\ D {\bf 18} (1978) 958; 
P.~F.~Harrison, D.~H.~Perkins and W.~G.~Scott, 
Phys.\ Lett.\ B {\bf 458} (1999) 79; 
Phys.\ Lett.\ B {\bf 530}, 167 (2002); 
Z.~Z.~Xing,
Phys.\ Lett.\ B {\bf 533} (2002) 85; 
X.~G.~He and A.~Zee,
Phys.\ Lett.\ B {\bf 560}, 87 (2003); 
\bibitem{rad}S.~Antusch, {\it et al.}, hep-ph/0305273.
\bibitem{GENIUS}H.~V.~Klapdor-Kleingrothaus {\it et al.}  
[GENIUS Collaboration], hep-ph/9910205.
\bibitem{cancel}E.g., 
S.~M.~Bilenky, S.~Pascoli and S.~T.~Petcov, 
Phys.\ Rev.\ D {\bf 64}, 053010 (2001); 
Z.~Z.~Xing, hep-ph/0305195. 
\end{thebibliography}
\end{document}